\documentstyle[11pt,epsfig,psfrag,wrapfig]{article}
\newcommand{\be}{\begin{equation}}
\newcommand{\ee}{\end{equation}}
\newcommand{\la}{\langle}
\newcommand{\ra}{\rangle}
\begin{document}
%
%
\title{ \Large
	Erratum to \\
	"Azimuthal asymmetry in electro-production \\
	of neutral pions in semi-inclusive DIS" \\
 	published in Phys.\ Lett.\ B {\bf 522} (2001) 37
	}
\author{A.~V.~Efremov$^a$, K.~Goeke$^b$, P.~Schweitzer$^{c}$ \\  
	\vspace{-0.1cm}\footnotesize\it
	$^a$ Joint Institute for Nuclear Research, Dubna, 141980 Russia\\
	\vspace{-0.1cm}\footnotesize\it
	$^b$ Institute for Theoretical Physics II, Ruhr University Bochum, 
	Germany\\
	\vspace{-0.1cm} \footnotesize\it 
	$^c$ Dipartimento di Fisica Nucleare e Teorica, 
        Universit\`a degli Studi di Pavia, Pavia, Italy} \date{} \maketitle
%
%
\vspace{-8cm}\begin{flushright} RUB/TP2-08/01E\end{flushright}\vspace{7cm}
%
%
Recently it became clear that the expression Eq.(115) in the paper
\cite{Mulders:1996dh} for the description of azimuthal $\sin\phi$
spin asymmetries in semi-inclusive hadroproduction in DIS on longitudinally
(with respect to the lepton momentum) polarized target contains a    
misprint in sign of the twist 3 term. This sign was corrected later 
in the paper \cite{Boglione:2000jk} (Eq.(2)).  However, all authors 
\cite{Oganessian:2000um,Ma:2001ie,Efremov:2001cz,Efremov:2000za} 
(including us) aiming at describing these phenomena did not notice this
very important change and, as a result, use the same sign for 
longitudinal (with respect to the virtual photon momentum) contribution
as for the transversal one. With the correct sign in Eq.(115) of 
ref.\cite{Mulders:1996dh} these 
contributions obtain opposite signs with positive sign for the longitudinal 
part if the z-axis is chosen in the direction of the virtual photon and 
positive target polarization is defined opposite to this direction, see Fig.1.
So all these descriptions should be recalculated with possibly different 
parameters for the Collins fragmentation function $H_1^\perp$.
%
%
\begin{figure}[b!]
\begin{tabular}{ccc}
	\includegraphics[width=7cm,height=3.5cm]{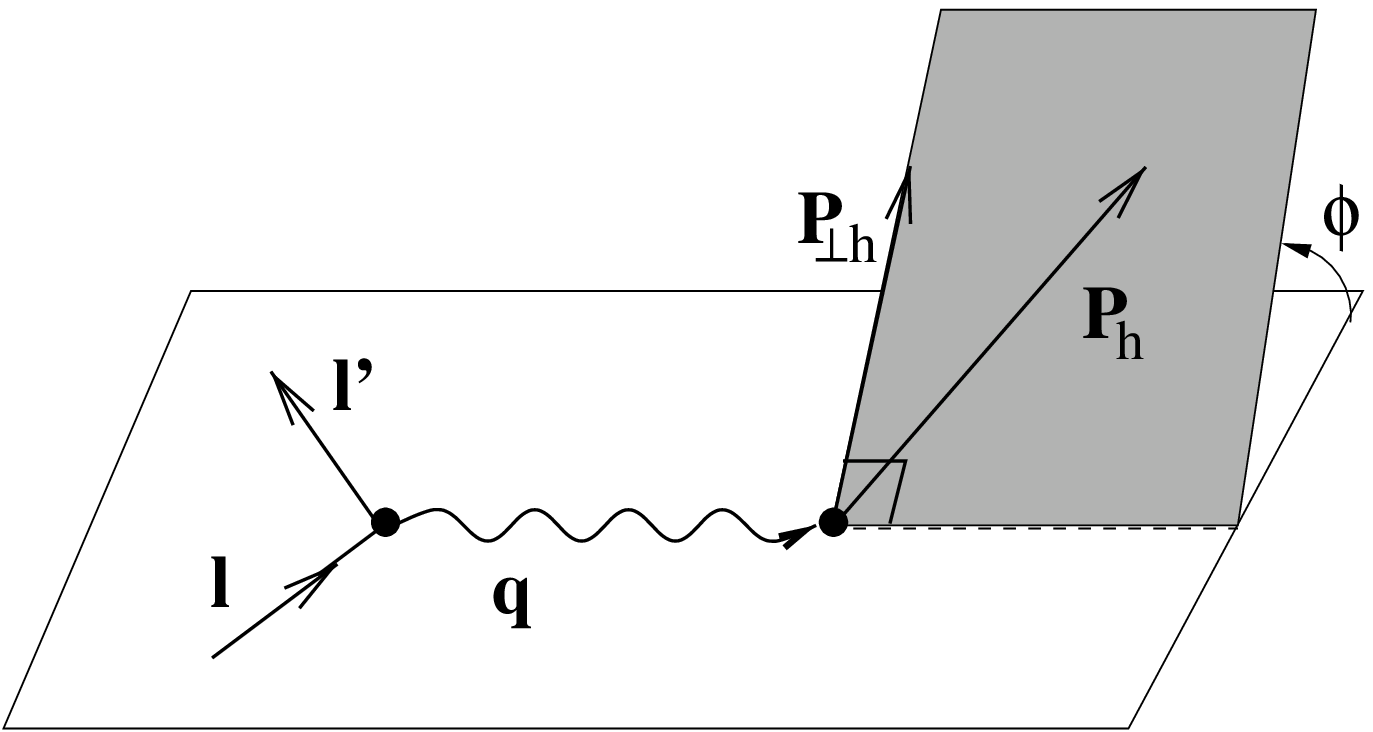} & &
	\includegraphics[width=4.5cm,height=4.5cm]{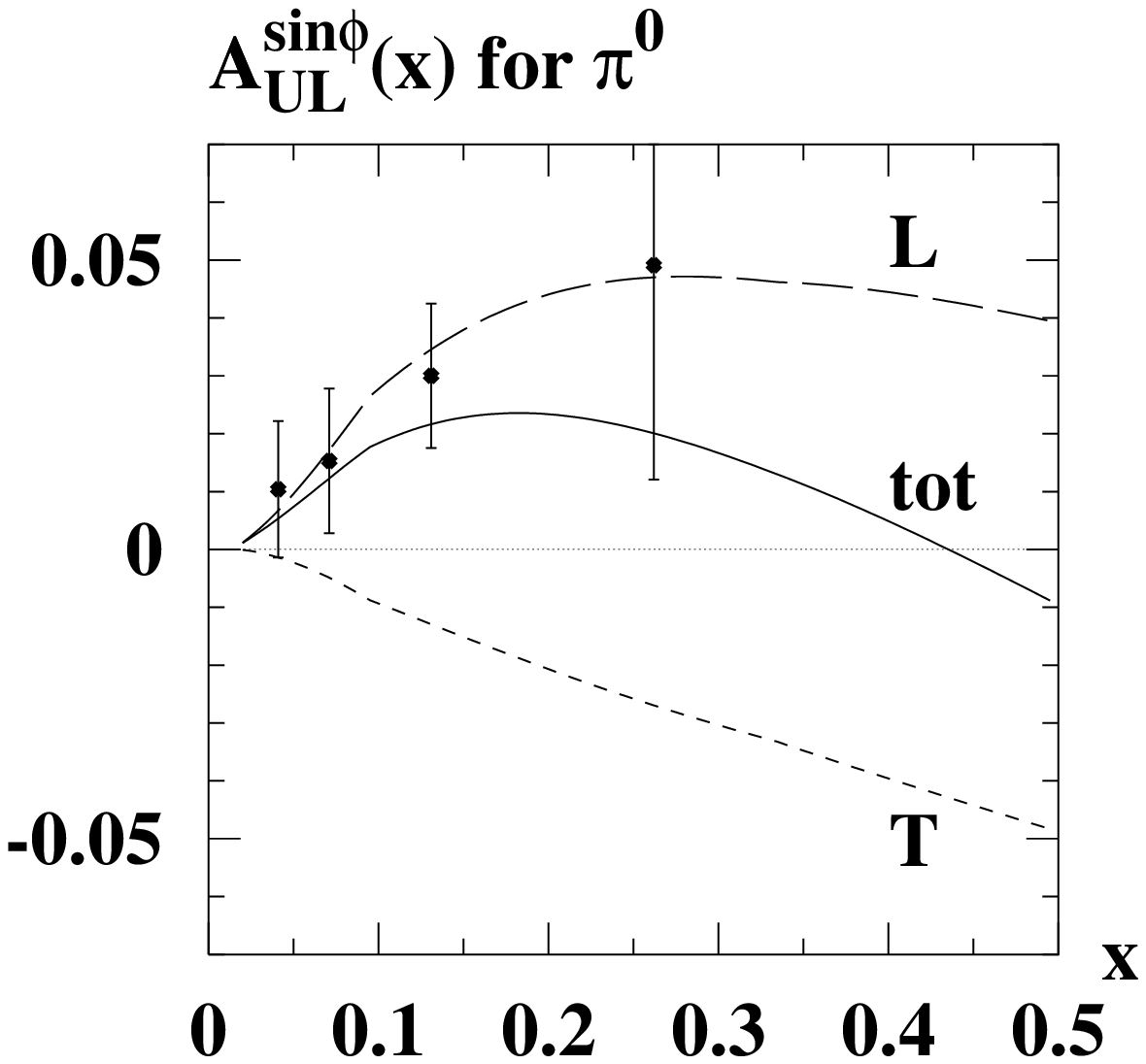}   \cr
    {\begin{minipage}[b]{2.9in}{
	\caption{
	{\footnotesize\bf Correction to Fig.2 in Ref.\cite{Efremov:2001cz}.}
	{\footnotesize\sl
	Kinematics of the process $lp\to l'hX$.
	Note the orientation of the azimuthal angle $\phi$ which corresponds
	to the convention of HERMES \cite{hermes}.
	In Refs.\cite{Mulders:1996dh,Boglione:2000jk} the azimuthal angle
	is defined as $(2\pi-\phi)$.}}}
     \end{minipage}} & $\phantom{XX}$ &
    {\begin{minipage}[b]{2.9in}{
    	\caption{
    	{\footnotesize\bf Correction to Fig.3c in Ref.\cite{Efremov:2001cz}.}
    	{\footnotesize\sl
    	The contribution of longitudinal (L, dashed) and
    	transverse (T, dotted) spin part to the total
    	(tot, solid line) azimuthal $\pi^0$ asymmetry
    	$A^{\sin\phi}_{UL}(x)$ and data from \cite{hermes-pi0} vs. $x$.}}}
     \end{minipage}}
\end{tabular}
\end{figure}
%
%

Concerning our paper \cite{Efremov:2001cz}, the azimuthal angle $\phi$ should 
be replaced by $(-\phi)$, see caption of Fig.1.
Due to this the expression for $\sigma_{UT}$ in Eq.(7), 
$B_T$ in Eq.(11) and in Fig.3a should have a minus sign  
(and similar changes in \cite{Efremov:2000za}).
With these changes and with using for Collins analyzing power the so called
"most reliable" value $\left|{\la H_1^{\perp}\ra\over\la D_1\ra}\right|
=(6.3\pm2.0)\%$  of DELPHI \cite{todd}, such recalculation results in 
asymmetry values about twice smaller than the experimental data. A better 
agreement is, however, achieved with the "optimistic" value of DELPHI 
\be\label{apower}
	\left|{\la H_1^{\perp}\ra\over\la D_1\ra}\right| =(12.5\pm 1.4)\%
\ee
obtained from the whole available interval of polar angles 
$15^\circ<\theta<165^\circ$ in the DELPHI experiment \cite{todd}. 
The results of these recalculations in comparison with the HERMES data 
are presented in Fig.2 and Fig.3 which replace Fig.3c and Fig.4 of 
Ref.\cite{Efremov:2001cz}.
%
%
\begin{figure}[t!]
\begin{tabular}{ccc}
	&&$\phantom{X}$ \cr &&$\phantom{X}$ \cr
	\includegraphics[width=4.5cm,height=4.5cm]{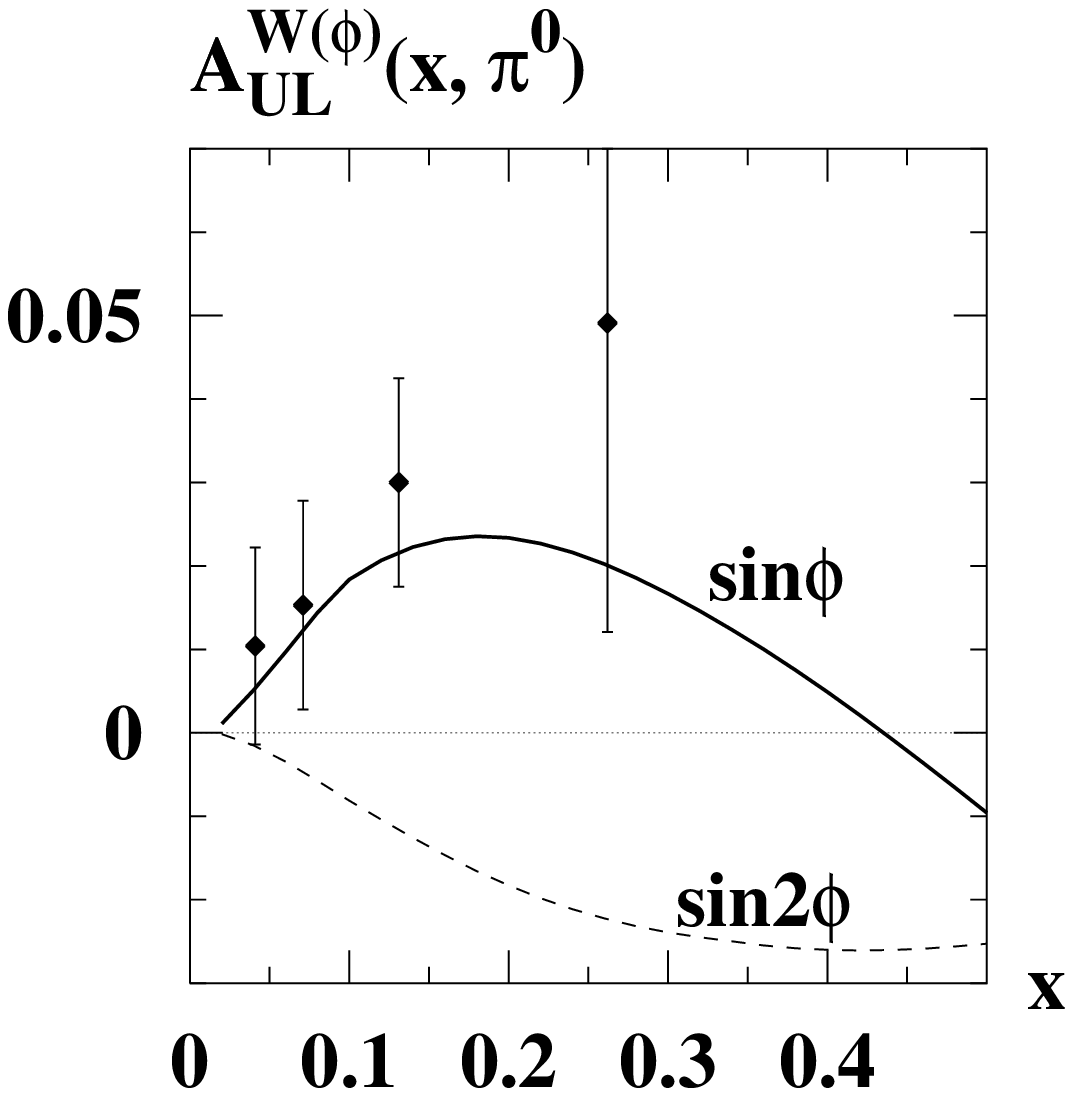} &
	\includegraphics[width=4.5cm,height=4.5cm]{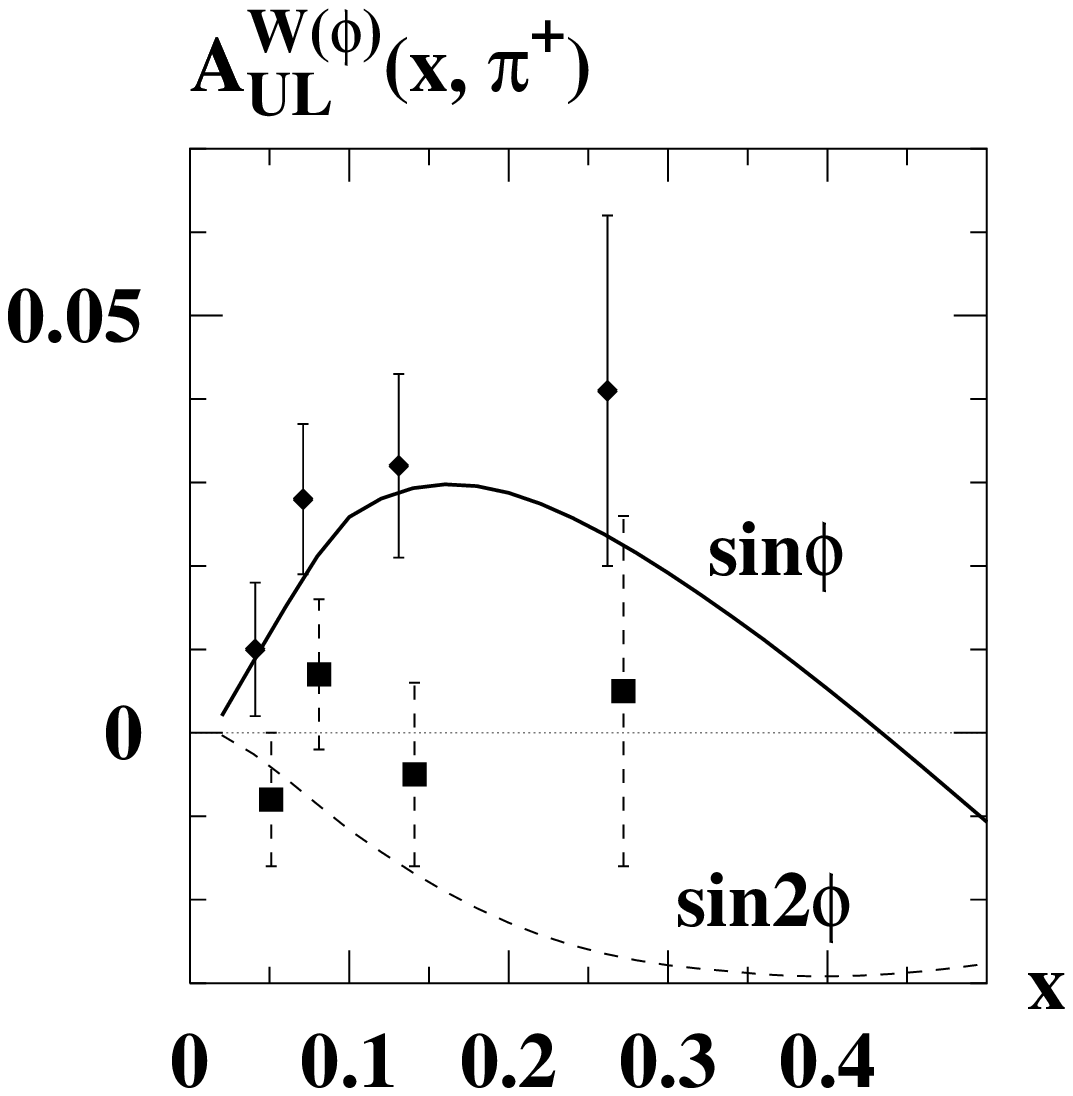} &
	\includegraphics[width=4.5cm,height=4.5cm]{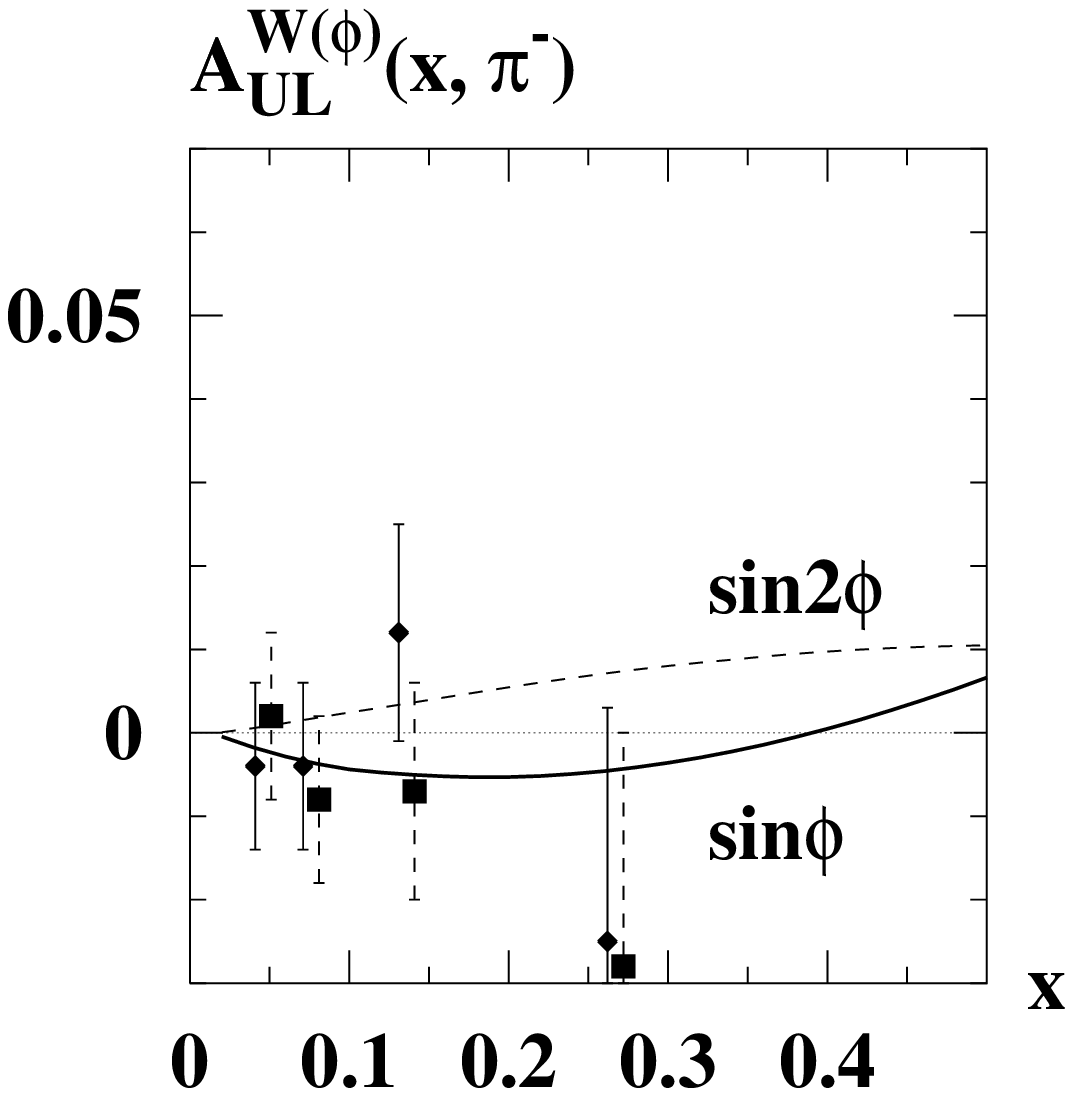} \cr
   {\footnotesize\bf Correction to Fig.4a in \cite{Efremov:2001cz}} &
   {\footnotesize\bf Correction to Fig.4b in \cite{Efremov:2001cz}} &
   {\footnotesize\bf Correction to Fig.4c in \cite{Efremov:2001cz}} 
\end{tabular}
	\caption{{\footnotesize\bf 
	Corrections to Figs.4a, 4b and 4c in Ref.\cite{Efremov:2001cz}}.
	{\footnotesize\sl
	Azimuthal asymmetries $A_{UL}^{W(\phi)}(x,\pi)$
 	weighted by $W(\phi)=\sin\phi$ (solid line) and 
	$\sin 2\phi$ (dashed line) 
	for the production of $\pi^0$, $\pi^+$ and $\pi^-$ as function of $x$. 
        The experimental data are from Refs. \cite{hermes-pi0,hermes}.
	Rhombs (squares) denote data for $A_{UL}^{\sin\phi}$
        ($A_{UL}^{\sin2\phi}$).
	The theoretical curves have an uncertainty due to the statistical and
	systematical error of the DELPHI result, eq.(\ref{apower}), and the
	theoretical uncertainty of the model.}}
\end{figure}
%
%

It is interesting to note that the negative sign of the transversal
contribution leads to a change of sign of asymmetries for $x>0.4$. 
This is due to a harder behaviour of $h_1(x)$ with respect to $h_L(x)$ 
(as seen in Fig.3b of ref.\cite{Efremov:2001cz}).
It should be noted that the prediction of $A_{UL}^{\sin\phi}(x,\pi)=0$
at $x\simeq(0.4-0.5)$ is sensitive to the approximation of favoured 
flavour fragmentation, which has been used in Ref.\cite{Efremov:2001cz}.
In principle one could conclude from data, how well this approximation
works. However, the upper $x$-cut is $x<0.4$ in the HERMES experiment 
\cite{hermes-pi0,hermes}.

The corrected values for the totally integrated asymmetries are 
\be\label{AUL-integrated}
	A_{UL}^{\sin\phi} = \cases{
	  \phantom{-}0.015 & for $\pi^0$ \cr
          \phantom{-}0.021 & for $\pi^+$ \cr
                   - 0.003 & for $\pi^-$} 
	\;\;\;\;\;\mbox{and}\;\;\;\;\;	
	A_{UL}^{\sin2\phi} = \cases{
	  \phantom{-}0.009 & for $\pi^0$ \cr
          \phantom{-}0.012 & for $\pi^+$ \cr
                   - 0.002 & for $\pi^-$}\ee
and replace the numbers in Table 1 of Ref.\cite{Efremov:2001cz}.
The numbers in Eq.(\ref{AUL-integrated}) have an uncertainty
due to the statistical and systematic error of the DELPHI result,
Eq.(\ref{apower}), and moreover an uncertainty of around $20\%$
due to the theoretical uncertainty of results from the chiral
quark soliton model.

The new estimate of the $z$-dependence of the analyzing power
$H_1^\perp(z)/D_1(z)$ from the $z$-behaviour of experimental asymmetries,
using as an input the transversities from  the chiral-quark soliton model
\cite{h1-model}, is presented at Fig.4 with a linear fit 
$$
H_1^\perp(z) = (0.33 \pm  0.06)\,z\,D_1(z)
$$
and with average ${\la H_1^\perp\ra}/{\la D_1\ra} = (13.8\pm 2.8)\%$
which is in good agreement with DELPHI result eq.(\ref{apower}).
\begin{figure}[ht!]
\vspace{-10mm}
\begin{center}
\begin{tabular}{ccc}
	\psfrag{H(z)/D(z)}{\boldmath$H_1^\perp(z)/D_1(z)$}
	\psfrag{z}{\boldmath$z$}
	\includegraphics[width=5cm,height=5cm]{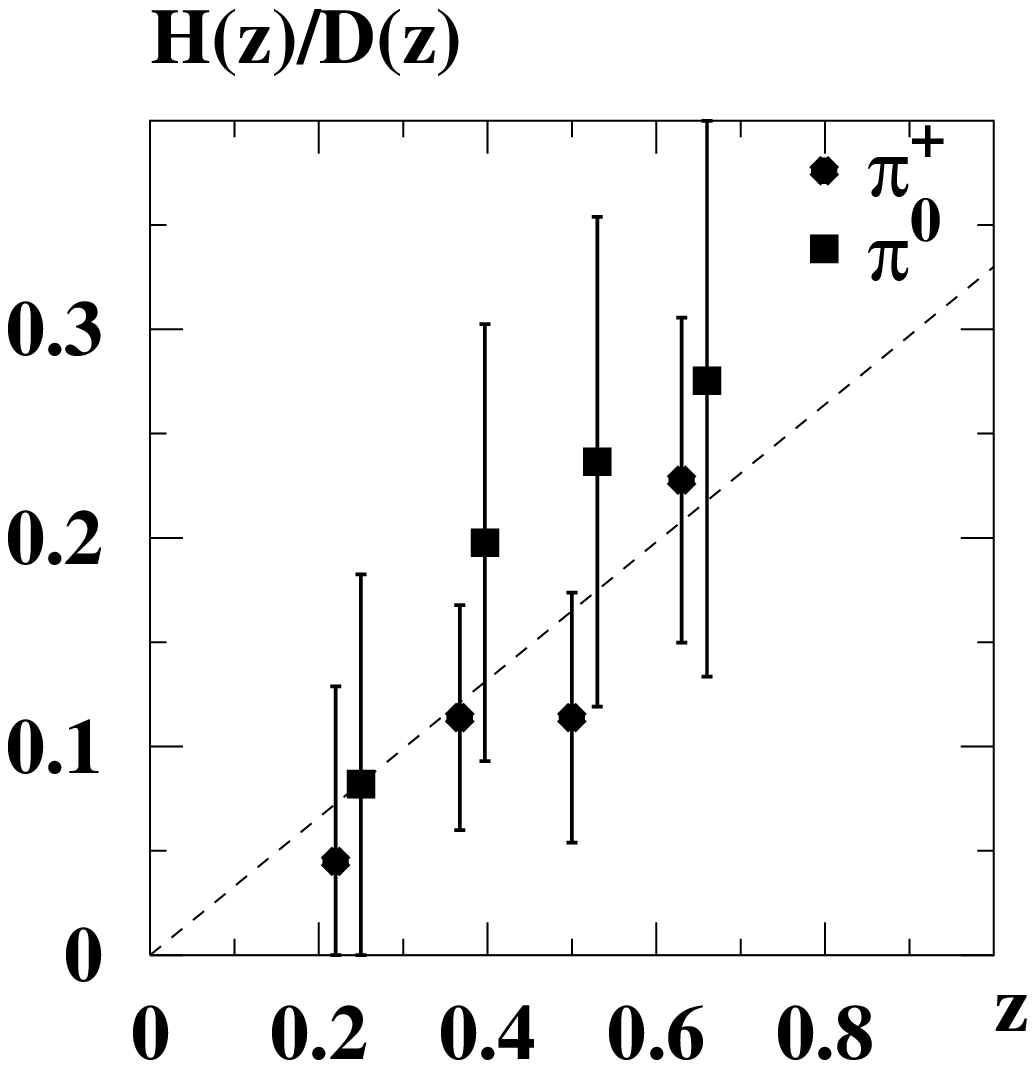}
& 	\hspace{-1cm} &
	\psfrag{H(z)/D(z)}{\boldmath$H_1^\perp(z)/D_1(z)$}
	\psfrag{z}{\boldmath$z$}	
	\includegraphics[width=5cm,height=5cm]{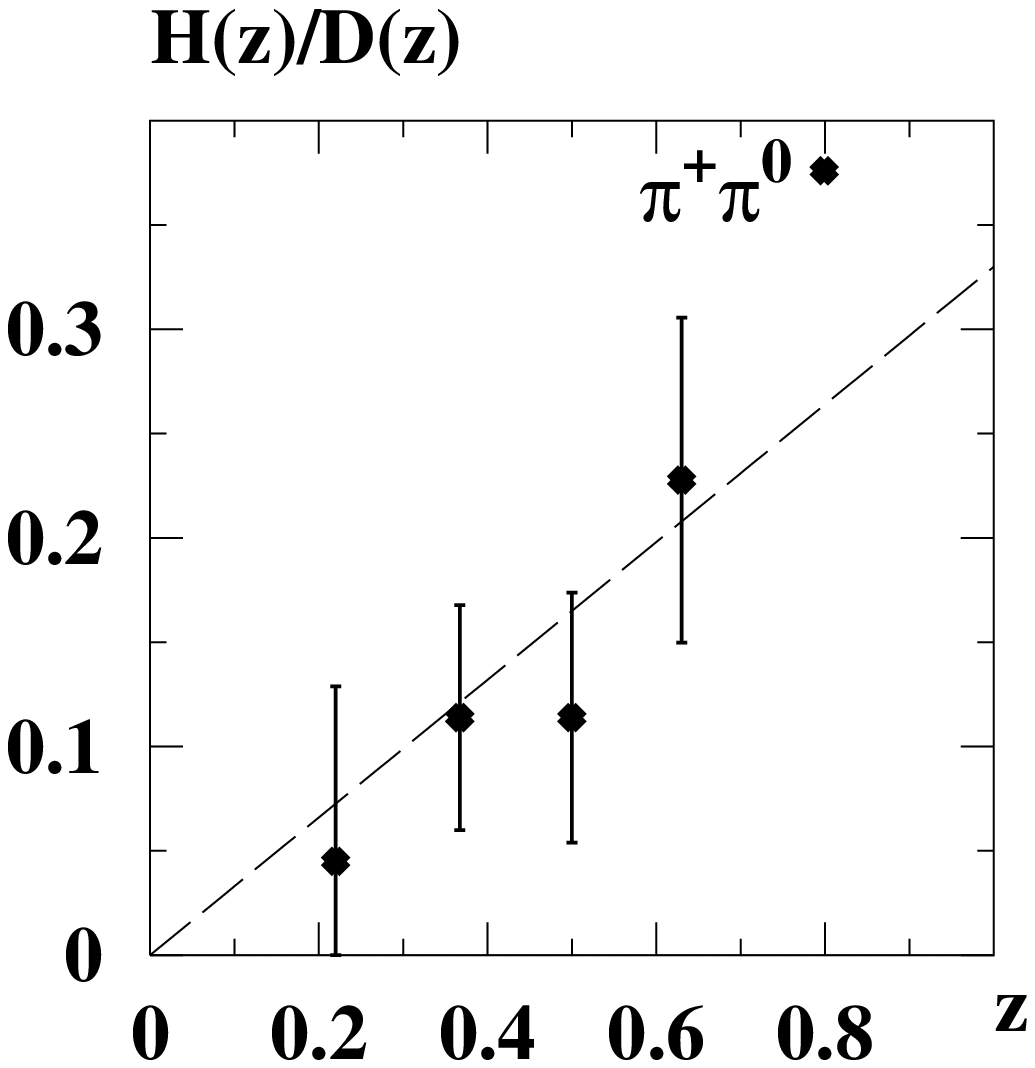}
\\ 
{\bf a} & \hspace{-1cm} & {\bf b} \end{tabular}
\end{center}
\vspace{-5mm}
	\caption{
	{\footnotesize\bf Corrections to Fig.5 in Ref.\cite{Efremov:2001cz}.}
	{\bf a.} 
	{\footnotesize\sl  $H_1^\perp(z)/D_1^\perp(z)$
	vs. $z$, as extracted from HERMES data
	\cite{hermes-pi0,hermes} on
	the azimuthal asymmetries $A_{UL}^{\sin\phi}(z)$ for $\pi^+$ and
	$\pi^0$ production using the prediction of the chiral quark-soliton
	model for $h_1^a(x)$ \cite{h1-model}. The error-bars are due to the
	statistical error of the data.}\newline
	{\bf b.} {\footnotesize\sl
        The same as Fig.4a with data points from $\pi^+$ and $\pi^0$ combined.
	The dashed line in both figures is the best fit
	to the form $H_1^\perp(z)/D_1^\perp(z)= a\,z$ with $a=0.33$.}}
\end{figure}
%
%

\vspace{0.2cm}

{\footnotesize \noindent
We would like to thank H.~Avakian and A.~Kotzinian for stimulating 
discussions, and P.~J.~Mulders for conversations on signs in
\cite{Mulders:1996dh,Boglione:2000jk}. 
The work of A.~E. is partially supported by RFBR grant 00-02-16696, 
INTAS grant 01-587, by Heisenberg-Landau Program and by BMBF and DFG. 
This work has partly been performed under the contract  
HPRN-CT-2000-00130 of the European Commission.}


\begin{thebibliography}{99}

\bibitem{Mulders:1996dh}
   	P.~J.~Mulders and R.~D.~Tangerman, 
	Nucl.\ Phys.\ B {\bf 461} (1996) 197, [arXiv:hep-ph/9510301] \newline
   	[Erratum-ibid.\ B {\bf 484} (1996) 538].
	

\bibitem{Boglione:2000jk}
	M.~Boglione and P.~J.~Mulders,
	Phys.\ Lett.\ B {\bf 478}, 114 (2000), [arXiv:hep-ph/0001196].

\bibitem{Oganessian:2000um}
	K.~A.~Oganessian, N.~Bianchi, E.~De Sanctis and W.~D.~Nowak,
	Nucl.\ Phys.\ A {\bf 689}, 784 (2001), [arXiv:hep-ph/0010261].
	\newline
	K.~A.~Oganessian, H.~R.~Avakian, N.~Bianchi and A.~M.~Kotzinian,
	arXiv:hep-ph/9808368.

\bibitem{Ma:2001ie}
	B.~Q.~Ma, I.~Schmidt and J.~J.~Yang,
	Phys.\ Rev.\ D {\bf 65}, 034010 (2002), [arXiv:hep-ph/0110324].

\bibitem{Efremov:2001cz}
	A.~V.~Efremov, K.~Goeke and P.~Schweitzer,
	Phys.\ Lett.\ B {\bf 522}, 37 (2001),
	\newline [arXiv:hep-ph/0108213].

\bibitem{Efremov:2000za}
	A.~V.~Efremov, K.~Goeke, M.~V.~Polyakov and D.~Urbano,
	Phys.\ Lett.\ B {\bf 478}, 94 (2000), [arXiv:hep-ph/0001119].

\bibitem{todd}
   	A.~V.~Efremov, O.~G.~Smirnova and L.~G.~Tkatchev,
   	Nucl. Phys. (Proc. Suppl.) {\bf74} (1999) 49 and {\bf79} (1999) 554; 
	[arXiv:hep-ph/9812522].

\bibitem{hermes-pi0}
	A.~Airapetian {\it et al.}  [HERMES Collaboration],
	Phys.\ Rev.\ D {\bf 64}, 097101 (2001), 
	[arXiv:hep-ex/0104005].

\bibitem{hermes}
  	A.~Airapetian {\it et. al.}, Phys.\ Rev.\ Lett.\  {\bf 84} (2000) 4047,
  	[arXiv:hep-ex/9910062]. \\
  	H.~Avakian, Nucl. Phys. (Proc. Suppl.) {\bf B79} (1999) 523.

\bibitem{h1-model}
   	P.~V.~Pobylitsa and M.~V.~ Polyakov, 
	Phys. Lett. {\bf B 389} (1996) 350 [arXiv:hep-ph/9608434].
   	\newline
   	P.~Schweitzer, D.~Urbano, M.~V.~Polyakov, C.~Weiss,
   	P.~V.~Pobylitsa, K.~Goeke, Phys. Rev. {\bf D 64}, 034013 (2001),
   	[arXiv:hep-ph/0101300].

\end{thebibliography}
\end{document}